\documentclass[aps,pra,groupedaddress,showpacs,onecolumn,sort&compress,floatfix, amssymb, noeprint]{revtex4-2}
\usepackage{float}
\usepackage[dvips]{graphicx}
\usepackage{dcolumn}
\usepackage{bm} 
\usepackage{amsmath,amsthm,amssymb}
\usepackage{graphicx}
\usepackage{color}
\usepackage{tabularx} 
\usepackage{booktabs} 
\usepackage[T1]{fontenc}
\usepackage[hidelinks]{hyperref}
\usepackage{upgreek}
\usepackage{xfrac}

\usepackage{units}				

\newcommand{\NA}{\text{NA}}
\newcommand{\SNR}{\text{SNR}}

\definecolor{gray}{rgb}{0.5, 0.5, 0.5}

\begin{document}
\title{A comparative study of deconvolution techniques for quantum-gas microscope images} 

\author{A.\ La Rooij\footnote{arthur.larooij@strath.ac.uk}, C.\ Ulm, E.\ Haller, S.\ Kuhr\ }
\address{University of Strathclyde, Department of Physics, SUPA, Glasgow G4 0NG, United Kingdom}

\date{\today}

\begin{abstract}
Quantum-gas microscopes are used to study ultracold atoms in optical lattices at the single-particle level. In these systems atoms are  localised on lattice sites with separations close to or below the diffraction limit. To determine the lattice occupation with high fidelity, a deconvolution of the images is often required. We compare three different techniques, a local iterative deconvolution algorithm, Wiener deconvolution and  the Lucy-Richardson algorithm, using simulated microscope images. We investigate how the reconstruction fidelity scales with varying signal-to-noise ratio, lattice filling fraction,  varying fluorescence levels per atom, and  imaging resolution. The results of this study identify the limits of singe-atom detection and provide quantitative fidelities which are applicable for different atomic species and quantum-gas microscope setups.\\

\end{abstract}

\maketitle
\section{Introduction}
Quantum-gas microscopes have revolutionized the study of ultracold atoms in optical lattices by allowing the simulation and observation of many-body quantum systems at the single-atom level \cite{bakr2009quantum,bloch2008many}. In recent years, researchers have been able to experimentally observe strongly correlated quantum phases, such as bosonic and fermionic Mott insulators \cite{sherson2010single,bakr2010probing,cheuk2015quantum,greif2016site,edge2015imaging}, anti-ferromagnetic phases \cite{boll2016spin,mazurenko2017cold, brown2017spin,  cheuk2016observation} and use ultracold gases to probe complex quantum dynamics \cite{bloch2008many,simon2011quantum,fukuhara2013quantum,preiss2015strongly,bohrdt2019classifying,kwon2022site} using direct fluorescence imaging of ultracold atoms. 
Many recent quantum-gas experiments use different atomic species in various lattice geometries \cite{yamamoto2020single,liu2021site,asteria2021quantum,nelson2007imaging,krahn2021erbium}.
In all these experiments, atoms are detected via fluorescence imaging, with a resolution close to the diffraction limit, which is on the order of the lattice spacing of about \unit[500]{nm} \cite{knottnerus2020microscope,picard2019deep,yamamoto2016ytterbium,phelps2019dipolar}. A high fidelity of the atom detection is essential for quantum simulation and quantum computation experiments, and an optimised and fast deconvolution can help to reduce the time required for imaging and identifying atoms.

However, determining  the presence or absence of an atom on a lattice site can be challenging in very dense atomic clouds, such as atoms in a Mott-insulating state with near unit filling, in which single empty lattices sites need to be identified with high fidelity \cite{fukuhara2013quantum}. In order to achieve this, the fluorescence images are usually processed using deconvolution techniques, such as the Lucy-Richardson (LR) method \cite{lucy1974iterative,richardson1972bayesian}, the Wiener deconvolution \cite{wiener1964extrapolation} and a local iterative (LI) deconvolution algorithm \cite{sherson2010single}. Which of these techniques is used varies throughout the community, and it is not a priory known which one achieves the highest atom detection fidelity in a specific experimental scenario. 
Recently, alternative approaches were introduced to reconstruct the atom distribution on a lattice using machine learning techniques \cite{picard2019deep} and a parametric deconvolution \cite{alberti2016super}.  
Other superresolution microscopy 
techniques have been invented  to resolve ultracold atoms in 1D trap geometries beyond the diffraction limit, by  probing them with a standing-wave, either in free space \cite{mcdonald2019superresolution, subhankar2019nanoscale}, or within a cavity \cite{deist2022superresolution}.

Here we present a systematic study of the Local Iterative, Lucy Richardson and Wiener  deconvolution methods  and characterise  their performance as a function of lattice filling and signal-to-noise ratio (SNR) of the images and specific details of the imaging system, such as magnification and pixelation. To characterise the the performance of the three methods when operating around and below the diffraction limit we, vary the lattice spacing for fixed microscope and imaging parameters. For our quantitative analysis, we created simulated images with a known atom occupancy that  match experimental images, and evaluated how  the deconvolution methods can retrieve the initial occupancy.

This article is structured as follows. In Section \ref{defSimulation} it is shown how we simulate realistic single-atom-resolved fluorescence images from a known atomic distribution on a square lattice, including a realistic noise model. In Section \ref{MethodsSec} we introduce the three deconvolution techniques used in this study. The results are presented in Section \ref{Results} where we first compare the different techniques using a uniform fluorescence signal from the atoms, then it is shown how the detection fidelity is affected by a varying fluorescence signal amplitude. In the last Section we vary the imaging resolution, magnification and camera pixel sizes to study how much signal is required to obtain an optimal reconstruction.

\section{Simulation of fluorescence images} \label{defSimulation}
In typical quantum-gas microscope experiments, atoms are detected in a two-dimensional optical lattice via fluorescence imaging by a high-numerical-aperture microscope objective. Localised on individual lattice sites, the fluorescing atoms can be approximated as ideal point sources. Depending on the numerical aperture \NA \ of the objective and the wavelength of the fluorescence light, $\lambda$, the image of a  point source is given by the peak-normalised point-spread function
\begin{equation} \label{eq:PSF}
\text{PSF}[\tilde{r}(x,y)]= \left[ \frac{2 J_{1}(\tilde{r})} {\tilde{r}} \right]^2 \text{, with    }   \tilde{r}(x,y) =  \frac{ 2 \pi\NA  }{\lambda}\sqrt{x^2 +y^2 },
\end{equation}
where $x$ and $y$ are the coordinates in the object plane and $J_{1}(r)$ is the  Bessel function of the first kind. 
The Rayleigh criterion states that in order to distinguish two point sources, the central intensity peak of one source must lie outwith the first minimum of the other, which is at $d_{\rm min}=0.61\lambda/\NA$. In an optical lattice of wavelength $\lambda_{\rm l}$, atoms in neighbouring lattice sites are separated by $a=\lambda_{\rm l}/2$, which is often of similar size compared to the diffraction limit $ d_{\rm min}$. As a result, it is challenging to resolve the atoms or to determine the presence of individual defects in densely filled  optical lattices, as it is the case, e.g., for atoms in a Mott-insulating state \cite{sherson2010single,bakr2010probing}.

It is the scope of this article to determine how accurately we can detect atoms in a fluorescence image as a function of signal-to-noise ratio, varying fluorescence levels, and of imaging resolution.  To study the effects of these parameters systematically, we create simulated images with a known atom occupancy, which we then retrieve using one of the deconvolution techniques mentioned above, giving access to a quantitative measure of the reconstruction fidelity. In an experimental realisation, a high-NA objective collects atomic fluorescence light, which is imaged onto a electron-multiplying CCD (EMCCD) camera. Noise in the images arises from  background counts and from the  electron-multiplying amplification process, which we both include in our simulations. Recently, also sCMOS cameras have been used for single-atom imaging. These cameras have a lower readout noise but also have a lower quantum efficiency \cite{kwon2022site}. As readout noise is not a dominant noise source, our simulations hold for both camera types. Other noise contributions such as  clock-induced charges are not simulated as they only have a marginal impact on the overall noise level \cite{alberti2016super}.   In the following we describe step by step how our simulated images are created. 

\begin{figure}[!htp] 
	\includegraphics[width=\textwidth ]{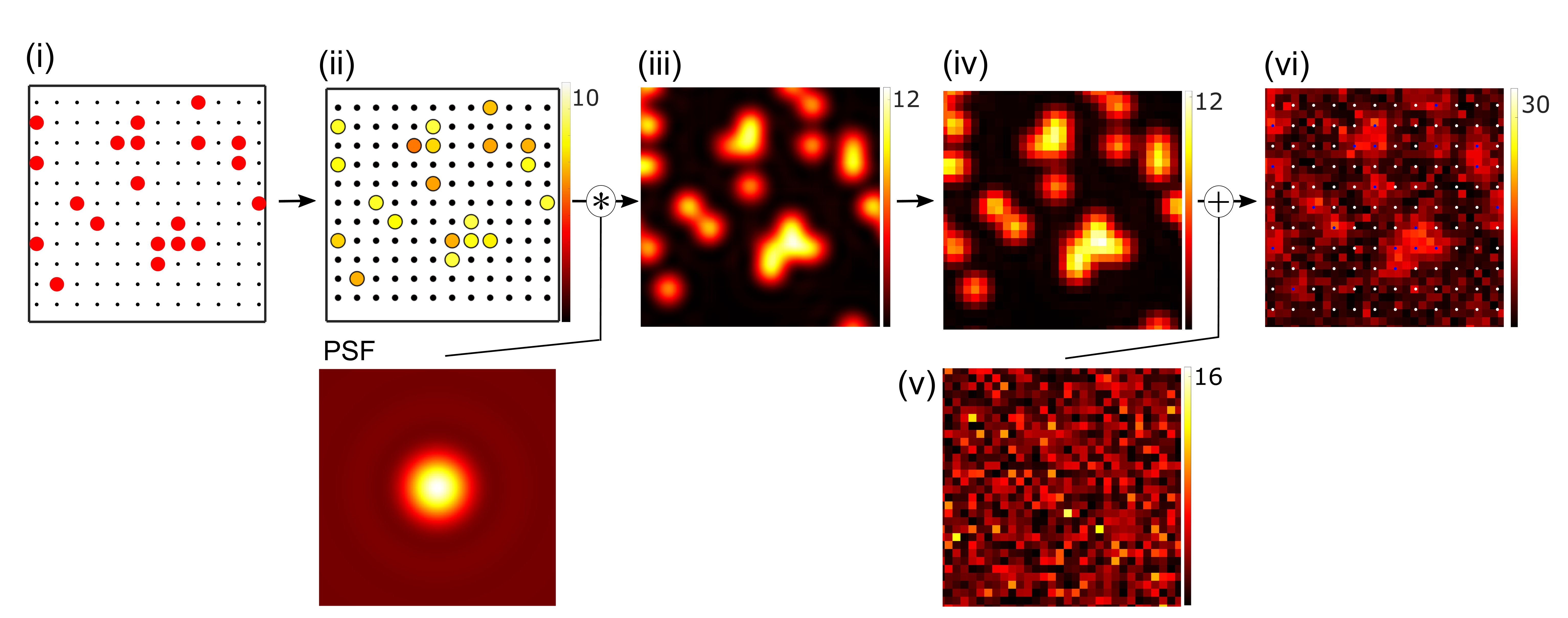}  
	\caption[ImagingProcess]{%
		\label{fig:Microscope} Simulation of fluorescence imaging of ultracold atoms in an optical lattice. The numbered panels refer to the simulation steps in the text. (i) the atomic distribution $n_{k,l}$,  (ii) the peak fluorescence intensity $F_{k,l}$ in the image plane, (iii) the  signal in the image plane $O(x,y)$ after convolution with the peak-normalised PSF shown below panel (ii), (iv) the signal as recorded by the camera showing pixelation and (v) the generated background noise. The final simulated image (vi) is the sum of (iv) and (v) including the Poissonian EMCCD gain. Each panel shows a region of $12\times 12$ lattice sites. The colorbars in panels (ii) and (iii) are each rescaled to match the scale of panel (iv), which is showing the counts per pixel ($\times 1000$). For the simulation shown above, we used $\sigma_{\text{BG}}=3750$ counts per pixel, SNR = 2 and $\Delta A/ \bar A =$ 40 \%. } 
\end{figure}

\begin{enumerate}
    \item We start with a known binary  occupation of atoms on the lattice, $n_{k,l}=0$ or $1$,  on a two-dimensional square array with $N^2 \approx 100\times100$ lattice sites with indices $k$ and $l$. The binary occupation reflects the fact that in most quantum-gas microscope setups, pairs of atoms on the same lattice site undergo light-assisted inelastic collisions and are lost before they emit any significant amount of fluorescence \cite{bakr2009quantum,sherson2010single}. We define as the filling factor, $\eta$, the fraction of lattice sites containing one atom.  The simulated images contain a central region of interest with $N^2$ lattice sites which are randomly populated with filling factors  $10 \% < \eta < 90 \%$. 

    \item  The lattice occupation $n_{k,l}$ is multiplied with the peak fluorescence intensity, $F_{k,l}$,  from  an atom on site ($k,l$). 
     
    \item The lattice occupation $n_{k,l}$  is convolved with the point-spread function PSF$(\tilde{r})$ of eq.~\ref{eq:PSF}, yielding the intensity distribution  $O(x,y)$ in the object plane:
    \begin{equation}\label{eq:image}
      O(x,y) = \sum_{k=0}^{N-1} \sum_{l=0}^{N-1}
      n_{k,l}F_{k,l} \text{PSF}[\tilde{r}(x- k a, y - l a]).
    \end{equation}
    
    \item  To obtain a simulated CCD camera image with image magnification $M$, the function $O(x/M,y/M)$ is discretized by integrating over the camera pixel size $d_{\text{pix}}$ (typically 5-20\,$\mu$m). This  yields the image, $I$, as a $512\times 512$ matrix. We define an amplitude, $A_{k,l}$, for each lattice site in the discretized image,  $A_{k,l} = \gamma F_{k,l}  d_{\text{pix}}^2 / M^2 $, 
    where the factor $\gamma$ takes into account detection efficiency and  EMCCD gain. If a PSF was centered on a pixel, the $A_{k,l}$ would correspond to the peak counts detected on the camera image. In the first part of our study, we have chosen the range of numerical values for the fluorescence intensity $F_{k,l}$, such that the pixel counts on the discretised image match our experimentally observed values \cite{haller2015single} for a specific pixel size (see further details in caption of Fig.~\ref{fig:Microscope}). 
    Initially, we use the  the same $A_{k,l}$ for each atom, and later simulate atoms with varying fluorescence, using a   mean amplitude, $\bar A$, and  variance, $\Delta A$.  
    We always keep the same number of pixels in the simulated images, and when we change the pixel size the amplitudes $A_{k,l}$ change accordingly, such that we keep the same  fluorescence intensity, $F_{k,l}$, per atom.
        
    \item  A random background signal $\Gamma_{\text{BG}}$ is added to each pixel of the simulated camera image. This background signal is dominated by stray photons from lasers used in the cooling and trapping process, which can vary significantly between experiments. The random values $\Gamma_{\text{BG}}$ have a half-normal probability distribution $P_{\text{BG}}(n|\mu, \sigma_{\text{BG}})$, with offset $\mu$ and width $\sigma_{\text{BG}}$:

\begin{equation}\label{eq:HalfNormal}
  P_{\text{BG}}(n|\mu, \sigma)= \sqrt{\frac{2}{\pi} } \frac{1}{\sigma_{\text{BG}}} \exp{\left[-0.5\left(\frac{n-\mu}{\sigma_{\text{BG} }}\right)^{2} \right]}; n \geq \mu 
\end{equation}
In our simulation we fix $\mu=200$ counts per pixel to match the offset count of our Andor Ixon DU-897 camera. The background noise level $\sigma_{\text{BG}}$ sets the counts on a single pixel \cite{haller2015single}. As $\sigma_{\text{BG}}$ represents noise originating from a region in the object plane, it changes when the pixelation or magnification is varied.  
This noise model accurately simulates our experimental images  \cite{haller2015single}. Using the the average,  $\bar{A}$, of the amplitudes $A_{k,l}$, we define the signal-to-noise ratio (SNR) of our simulated images as  $\text{SNR} = {\bar{A}}/{\sigma_{\text{BG}}}$. It should be noted at this point that our SNR does not depend on the pixel size $d_{\text{pix}}$ or magnification, as we consider that the predominant noise source originates from stray light in the object plane, that should scale the same way as the fluorescence amplitude from the atoms. We do not include pixelation or discretisation noise in our definition of the SNR. 
    
    \item 
      The amplification process of the EMCCD camera as well as the discretisation of the image due to the camera's pixel size introduce noise. This amplification noise is typically significantly smaller than the noise level from the background signal.  In our case, a Poissonian filter is applied to simulate the signal amplification process of the EMCCD camera \cite{andor,MatlabPoisson}. 
\end{enumerate}

Fig. \ref{fig:Microscope} illustrates the different steps involved in the simulations process. 
For the simulations in Sections \ref{Results}a-b  we have chosen $\lambda=770.1$\,nm, $\NA =0.69$, $a=532$\,nm, $M=72$, $\sigma_{\text{BG}}=3750$ counts per pixel and a pixel size of $d_{\text{pix}} = $ \unit[13]{$\mu$m}. These parameters match those used for fluorescence imaging of ultracold $^{40}$K atoms in a square lattice ($\lambda_{\rm l }=1064$\,nm) \cite{haller2015single}, with a diffraction limit $d_{\text{min}}=$ \unit[691]{nm}.  For these parameters, the atoms are spaced at $a=\lambda_{\rm l }/2$ and cannot be considered resolved using the Rayleigh criterion ($a = 0.88\,d_{\text{min}}$). To test the deconvolution methods, we generate different sets of simulated images in which we vary the SNR by changing $\bar{A}$, and later include $\Delta A$ as well. We always consider a square lattice in which we vary the lattice filling $\eta$. In the last section we also vary the lattice spacing and pixelation.   

Our goal is to identify the best deconvolution algorithm capable of reconstructing the known lattice occupation $n_{i,j}$  with the highest fidelity, first as a function of lattice occupation, $\eta$, and SNR only, then also including varying atomic fluorescence  levels and resolution.  While such a reconstruction is possible with high fidelity in the case of sparse filling, this will be increasingly difficult in the case of dense filling and low signal.

\section{Deconvolution techniques} \label{MethodsSec}
In this study, we use three different deconvolution methods, a 'local iterative' (LI) deconvolution algorithm, the Lucy-Richardson (LR) algorithm, and the Wiener deconvolution. Both Wiener and LR algorithms are widely used in microscopy applications ranging from biology to astronomy \cite{manwar2021signal, starck2002deconvolution}. 
The LI deconvolution algorithm was used in one of the first quantum-gas microscope experiments  \cite{sherson2010single} to determine the lattice occupancy of Mott-insulator images with dense filling. All our simulation and deconvolution methods were implemented in MATLAB \cite{WienerMatlab,LucyMatlab}. 
The algorithms use a discrete PSF as an input ($I_{\text{PSF}}$), which can be measured with a higher resolution than the microscope images by averaging multiple images of single atoms using sub-pixel scaling \cite{haller2015single}. In our study, a PSF with a resolution five times higher than the imaging  resolution is used.
In the following, we provide a detailed description of  each method. 

The LI algorithm \cite{sherson2010single,schauss2015high} finds the most likely distribution of atoms by placing point-spread functions of variable amplitudes on the known lattice grid, using an optimisation method similar to a fitting algorithm, minimising  the difference between the camera image and a reconstructed image. First, based on the local intensity profile and the known average fluorescence signal per atom, an initial guess of the lattice occupation is made. Based on this, the algorithm  optimises  sub-images of $10\times 10$  sites and creates different configurations of the lattice occupation of the central $3 \times 3$ sites, by, e.g., removing or adding an atom on a lattice site, or placing an atom on a neighbouring site. In a second step, the algorithm adjusts the amplitudes $\tilde{A}_{k,l}$ of the central site  and its nearest neighbours, by decreasing or increasing the amplitude on each lattice site in discrete steps of 3~\% of the maximum signal. After each change of the local configuration or amplitudes, a convolved sub-image image is generated, and subtracted from the corresponding camera sub-image. An error is calculated by summing over the squares of all pixel counts of the difference image, and if this error is smaller than the previously found optimum value, the new configuration is kept. Once all local configurations have been tested, the algorithm starts with the first  sub-image again and the same optimisation is repeated until the error is no longer reduced. 

\begin{figure} [!hbp]  
 	\includegraphics[width=0.95\textwidth]{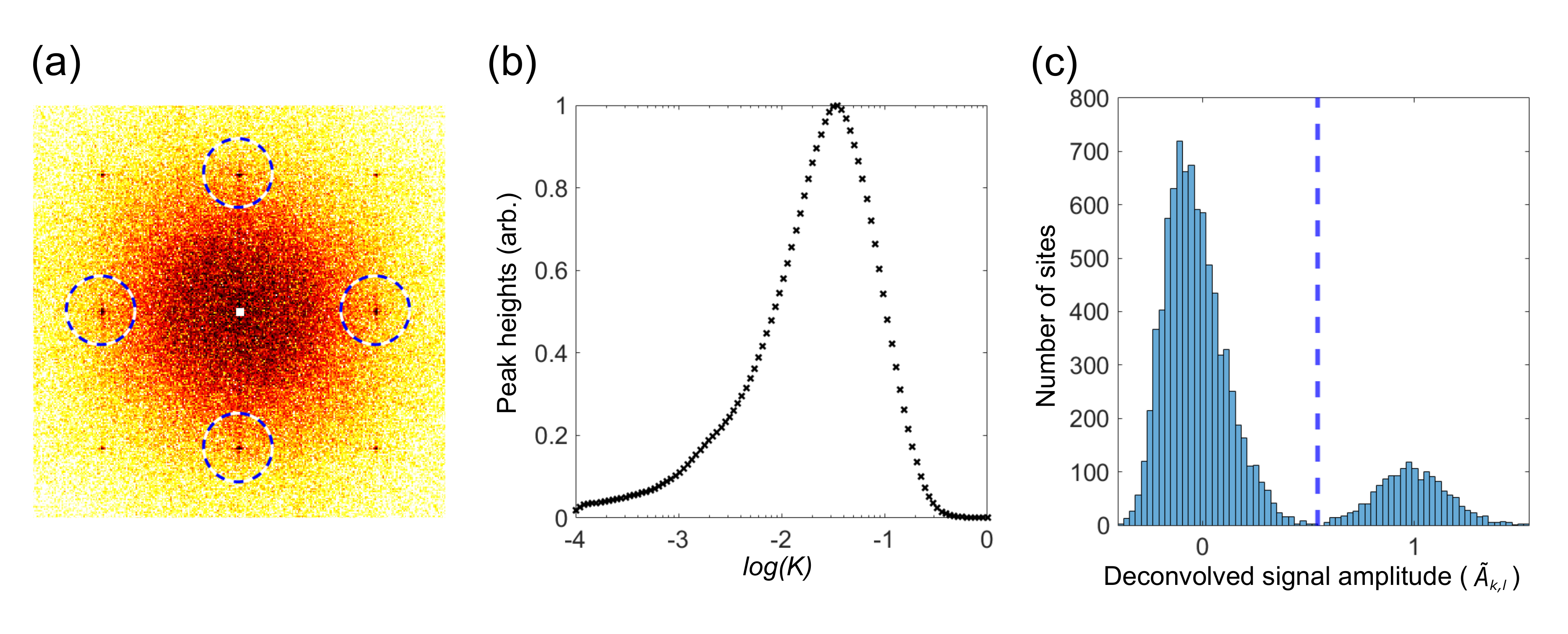} 
 	
 	\caption[AutoK]{%
		Optimising the Wiener deconvolution protocol. a) FFT of an image after Wiener deconvolution with the optimal deconvolution parameter $K$. The four encircled peaks are the first-order harmonics at the inverse lattice spacing (the zeroth order in the center is removed to enhance the contrast). The dashed circles are a guide to the reader only.  b) The signal at these four peaks as a function of  $K$ allows us to determine its optimal value. This plot is normalised. c) The histogram of deconvolved signal amplitude per site $\tilde{A}_{k,l}$ for using the optimum $K$ found in b). The threshold above which sites are considered occupied is shown by the dashed line. 
\label{fig:AutoK}
  } 
 \end{figure}

\begin{figure} [!htbp]  % I would like this at 
 	\includegraphics[width=0.99 \textwidth]{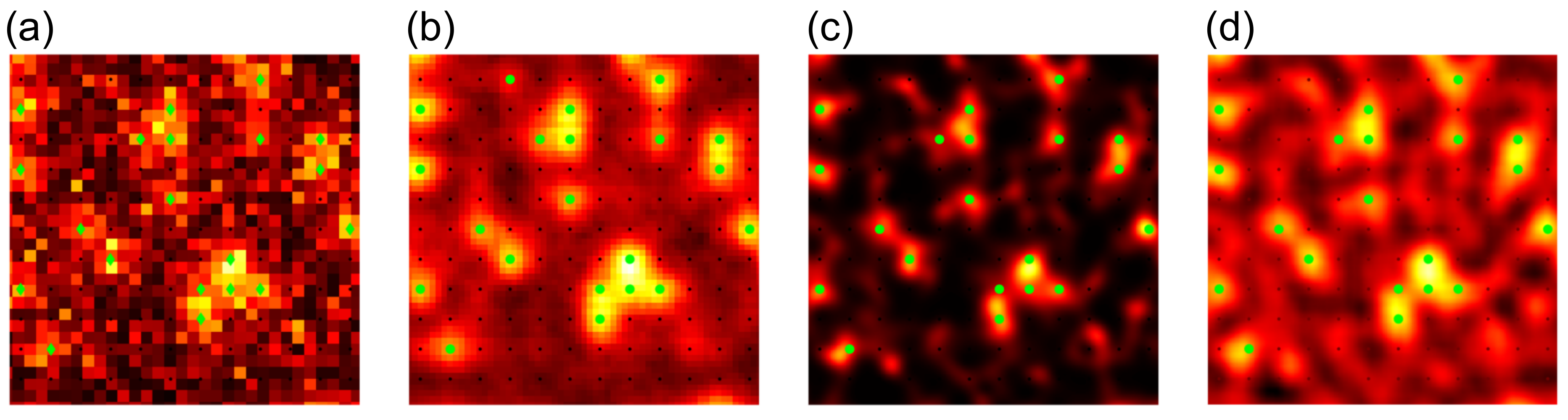} %  bb=0 0 500 500
 	\caption[Methods]{%
		\label{fig:Methods} Comparison of deconvolved images. a) Section of a simulated image with low signal-to-noise ratio ($\text{SNR}=2$, $\sigma_{\text{BG}}=3750$ counts per pixel, $\Delta A/ \bar A =$ 40 \% and a filling level of \unit[10]{\%}), generated using the protocol described in Section 2. Occupied sites are indicated by the green diamonds. b) Reconstructed image generated by the Local Iterative deconvolution algorithm.  c)  Deconvolved image created by the Lucy-Richardson algorithm, and d) by the Wiener method with optimised $K$ parameter. The sites identified as occupied by each method are marked with green circles and empty sites as small black dots. All images are normalised to allow for better comparison.  } 
 \end{figure}

A typical image after LI deconvolution is shown in Fig. \ref{fig:Methods}b for $\text{SNR}=2$. A histogram of the resulting amplitudes, $\tilde{A}_{k,l}$, shows two peaks, corresponding to empty and occupied lattice sites, as shown in Fig.~\ref{fig:AutoK}c). We fit the distribution with a bimodal Gaussian, and find a threshold value between the two peaks at the intersection of the two Gaussians. A lattice site with a signal amplitude above the threshold is considered occupied ($\tilde{n}_{k,l}=1$). Finally, we calculate the reconstruction fidelity, $F$, as the fraction of correctly identified lattice site occupations,
\begin{equation}
F= 1 - \frac{1}{N^2} \sum_{k,l=1}^{N}\left|n_{k,l} - \tilde{n}_{k,l}\right| \,.
\end{equation}

The second algorithm is the Wiener deconvolution \cite{wiener1964extrapolation}, a method that employs an inverse Fourier filter weighted by a factor $K$ that is proportional to the noise-to-signal ratio and includes the Fourier transform of both the camera image $I$ and the discretised PSF, $I_{\text{PSF}}$, 
\begin{equation}\label{eq:Wiener}
    W= {\mathcal{\widetilde F}} \left[ \frac{ \mathcal{F}(I) }{ \mathcal{F} (I_{\text{PSF}} )}   \frac{| \mathcal{F}(I_{\text{PSF}})|^{2}}{| \mathcal{F}(I_{\text{PSF}})|^{2} + K}   \right] \,,
\end{equation}
where $W$ is the deconvolved image, and $\mathcal{F}(.)$ and $\mathcal{\widetilde F}(.)$ are the Fourier and inverse Fourier transforms, respectively. The image $I$ needs to be upscaled without interpolation \cite{MatlabImresize} 
to the resolution of the PSF to correctly deconvolve.

As $K$ is a priori unknown, we developed a protocol to determine the optimum  $K$ value for a given image. The regular spacing of the atoms in the lattice leads to peaks in the fast Fourier transform of the deconvolved images, at a distance from the centre corresponding to the inverse lattice spacing [Fig.~\ref{fig:AutoK}a)]. We tested this method and found it works efficiently for all fillings over \unit[10]{\%}. To find the optimum $K$ for which these peaks are most prominent, we compare the peak heights for images deconvolved with different $K$ [Fig.~\ref{fig:AutoK}b)]. To reduce fringes occurring during the Fourier transforms, we blur the outer 30 pixels in each image \cite{WienerMatlab}. After deconvolution with the optimal $K$ value, the signal of each lattice site is integrated within a circle of radius $a/2$, yielding the signal amplitudes $\tilde{A}_{k,l}$. For the simulations used in figures 3-5 only \unit[66]{\%} of the signal is contained within this radius and it increases after the deconvolution process. As for the LI deconvolution, the histograms  of the $\tilde{A}_{k,l}$ show a bimodal distribution, corresponding to empty and occupied lattice sites [Fig.~\ref{fig:AutoK}c)], and again a threshold value is found by a bimodal Gaussian fit. An example of a Wiener-deconvolved image is shown in Fig. \ref{fig:Methods}d.

The Lucy-Richardson (LR) deconvolution is a Bayesian-based maximum-likelihood estimation, capable of calculating the most likely original image in the presence of noise and blurring due to convolution with a (known or unknown) point-spread function. We use the MATLAB function {\sc deconvlucy($I,I_{\text{PSF}},m $)}, with the input image $I$, the PSF, and the number of iterations $m$, which is a free parameter. We found that a fixed number of 7 iterations results in the best atom  reconstruction fidelity for all parameters regimes discussed in Sections \ref{Results}a-b. When performing more than the optimum number of iterations, the algorithm is known to start amplifying noise instead of the signal, which in our case results in a reduction of the reconstruction fidelity.  As for the Wiener deconvolution, the image $I$ needs to be upscaled without interpolation and the edges of the images are blurred. We also found that the fidelity is increased when we apply a low-pass Fourier filter to the original image because it reduces the effect of pixelation. The filter has a cutoff set by the spatial frequency of $1/(\unit[1.8]{ d_{min} } )$. After deconvolution, the signal around each lattice site is integrated to give the deconvolved amplitudes $\tilde{A}_{k,l} $ and the lattice occupation is calculated in the same way as for the Wiener method. 

We tested each deconvolution technique in combination with other filters, such as higher frequency low-pass filters, not filtering at all, and by using various weighting and masking options provided by MATLAB. The results below are generated by the methods described above and represent the most successful version of each deconvolution protocol. We found that the Wiener and LR methods work well with offset present, while for the LI method is it necessary to subtract the offset such that the average  background intensity is zero. To do this, we simulate a background image of which we calculate the mean pixel value  $(\approx \mu + \sigma_{\text{BG}})$. In many experiments it has been  customary to subtract not an average background but a single background image as a way to correct for variations of the background in time. This however increases the noise level, which can be avoided by subtracting an averaged background image, similar to the fringe-removal protocol used in absorption imaging \cite{ockeloen2010detection, niu2018optimized}. To  compare the  three deconvolution methods, we used the LI algorithm  with a subtracted averaged  background image, as we did for the LR and Wiener method.   

For all deconvolution methods, we upscaled without interpolation \cite{MatlabImresize} the resolution of the simulated images by a factor of five in order to match the resolution of the PSF. We can alternatively downscale the PSF to match the image resolution, which greatly increases the computation speed, but it reduces the atom detection fidelity by at least several percent.  
We qualitatively compare the three deconvolution methods and the resulting estimated occupation using a test image with $\text{SNR}=2$  (Fig.~\ref{fig:Methods}). Such a low SNR makes it challenging to identify atoms by eye, and we can see how each method increases the image contrast in a different  way.
The LR algorithm reduces the noise in the image while the Wiener method creates a wavelike pattern with high maxima at occupied sites. 
The LI method uses prior knowledge of the underlying optical lattice for the deconvolution. In a previous study \cite{alberti2016super}, a deconvolution technique has been introduced that can be seen as a combination of the LI and Wiener method. In such a 'parametric deconvolution', knowledge of the  experiment is used, such as lattice structure and specific noise properties. To our knowledge, this technique has not been  implemented in a 2D experiment to date. 

\section{Results} \label{Results}

\subsection{Comparing deconvolution methods}  
\begin{figure}[!ht] 
	\includegraphics[width=\textwidth ]{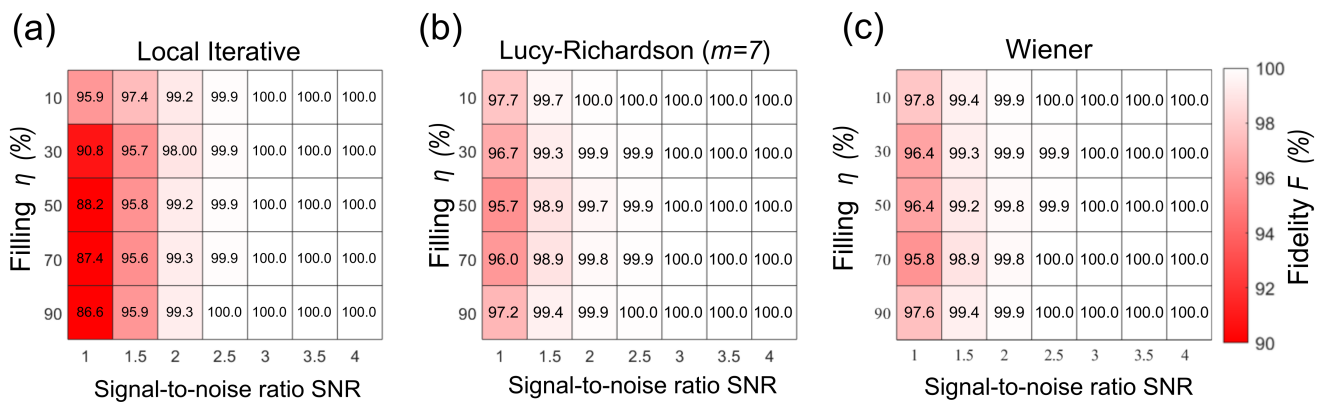} 
	\caption[MethodsRow_rev2_V04]{%
		\label{fig:MethodsComaprison} Comparison of deconvolution methods. Reconstruction fidelity  for simulated images with varying SNR and lattice filling fraction $\eta$, using a) the Local Iterative deconvolution,  b) the Lucy-Richardson algorithm (7~iterations), c) the Wiener deconvolution with optimised $K$ parameter.
		\label{fig:MethodsRow}
 } 
\end{figure}

We first compare the reconstruction fidelity of the different deconvolution methods (Fig.~\ref{fig:MethodsRow}) for varying lattice filling and signal-to-noise ratios, with the same simulated fluorescence signal for all atoms  ($\Delta A= 0$). We find that the LR and Wiener methods have the highest reconstruction fidelity, with  $F>95\%$ for $\SNR=1$ and $F>99\%$ for $\SNR=2$. In comparison, the LI method  only achieves $F<90\%$ for $\SNR=1$ for filling factors $\eta > 50\%$, and $F$ decreases further with increasing $\eta$. Both Wiener and LR methods are least reliable around 50\% filling, and require  $\SNR > 1.5$ to obtain an error rate of below 2\%, and both achieve $F>99.9\%$ for $\SNR>2.5$.  The fidelity of the LI method is less good throughout for all parameters and most noticeably for higher filling and $ \text{SNR}<2$. 

A key difference between the methods is the computation time required. On a standard desktop PC, the LI method requires up to two minutes to deconvolve an image and the computation time scales with the number of occupied sites. In contrast, the LR and Wiener methods (fixed $K$) require only several seconds per deconvolution, independent of $\eta$. The automated Wiener protocol can require up to one minute per image, as about $40$ deconvolutions are needed to find the best $K$.

\subsection{Including inhomogeneous fluorescence}
\begin{figure}[!h]
\begin{center}
\includegraphics[width=0.99 \textwidth ]{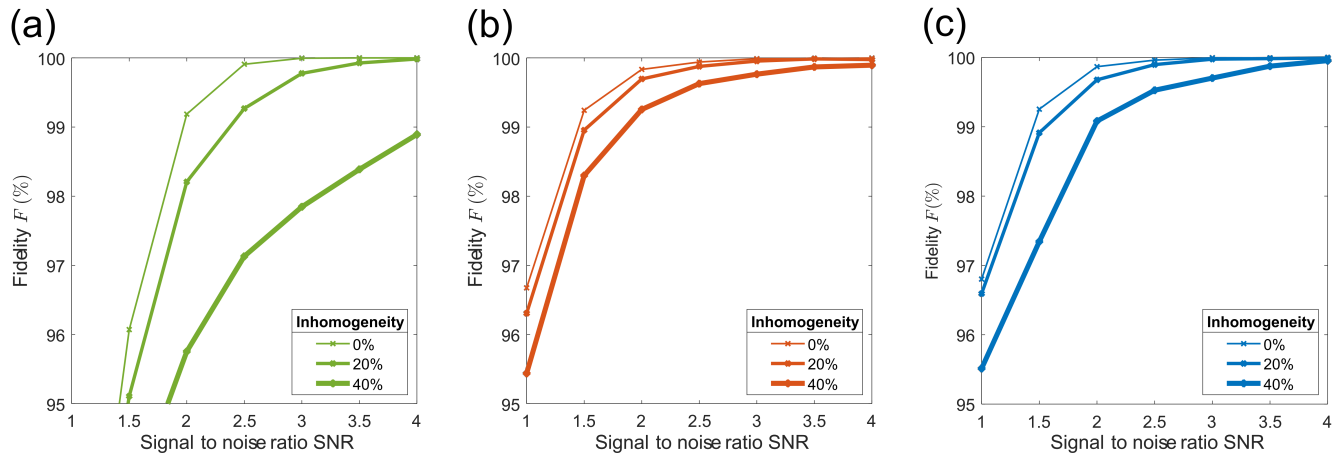} 
\caption[ANScaling]{%
	Atom-detection fidelity in the presence of inhomogeneous fluorescence with $\Delta A / {\bar A} = (0,  \pm20\%$, and $\pm40\%$) after deconvolution using a) the Local Iterative method, b) the Lucy-Richardson algorithm and c) the Wiener algorithm. The results shown are an average over the five filling levels (10\%, 30\%, 50\%, 70\% and 90\%) used in Fig~\ref{fig:MethodsComaprison}.    \label{fig:ANScaling}
 }
      \end{center}
\end{figure}

In real quantum-gas microscope images, the amount of fluorescence light emitted varies from atom to atom, e.g., due to inhomogeneous  cooling laser-light intensity or polarisation, or due to spatially varying trapping frequencies. Atoms also have a small but finite probability of being  heated out of a lattice site and being re-trapped on a different site. These loss and hopping events can also reduce the amount of fluorescence light collected from an atom.
While one could in principle simulate these processes individually, we choose to consider only the resulting effect of inhomogeneous fluorescence. For this purpose we simulated microscope images with a variation of $\pm20\%$ and $\pm40\%$ of the amplitudes $A_{i,j}$ around the mean value $\bar A$  ($\Delta A/\bar A=0.2$ and $0.4$). We again use the three different deconvolution algorithms and compare the reconstruction fidelity  (Fig.~\ref{fig:ANScaling}) after averaging over the five filling levels used previously. As expected, for all methods we see a  drop in detection fidelity for increasing inhomogeneity. We also find that the Wiener and LR methods are less prone to inhomogeneous fluorescence ($F>99.5\%$ for $\SNR=2.5$, $\Delta A/\bar A = 0.4$), compared to the LI algorithm for which we see a lower fidelity ($F<97\%$) using the same parameters.

\subsection{Varying the resolution}
We now consider the effect of the image magnification and pixelation on the reconstruction fidelity. In many quantum-gas microscope experiments, one lattice site is imaged to about $4\times4$ pixels \cite{sherson2010single,bakr2010probing,cheuk2015quantum,greif2016site,edge2015imaging,yamamoto2020single,liu2021site,asteria2021quantum,nelson2007imaging,krahn2021erbium, knottnerus2020microscope,picard2019deep,yamamoto2016ytterbium,phelps2019dipolar}.  Several groups have  shown how sub-wavelength traps can be created for ultracold atoms 
\cite{la2018deposition, gonzalez2015subwavelength, wang2018dark}, and in a recent study, Erbium atoms in an optical lattice with a spacing of $a =266$\,nm are deemed resolvable by quantum-gas microscopes \cite{picard2019deep}. Single-atom detection  in such systems is  challenging and our simulations will help to understand how SNR and resolution affect the deconvolution methods. A microscope-imaging setup can be characterized by the objective's numerical aperture, $\NA$, the fluorescence wavelength, $\lambda$, and the lattice spacing, $a$. The magnification, $M$, and pixelation, $p$ ($p= M a / d_{\text{pix}}$), can be varied independently of this. In previous studies,  the minimum number of pixels required to resolve atomic positions on a lattice without loss of information is reasoned to be $p=2.5$ \cite{pawley2006points,alberti2016super}, based on linear-optics calculations (Shannon-Nyquist sampling). 

\begin{figure}[!h] 
	\includegraphics[width=0.95\textwidth ]{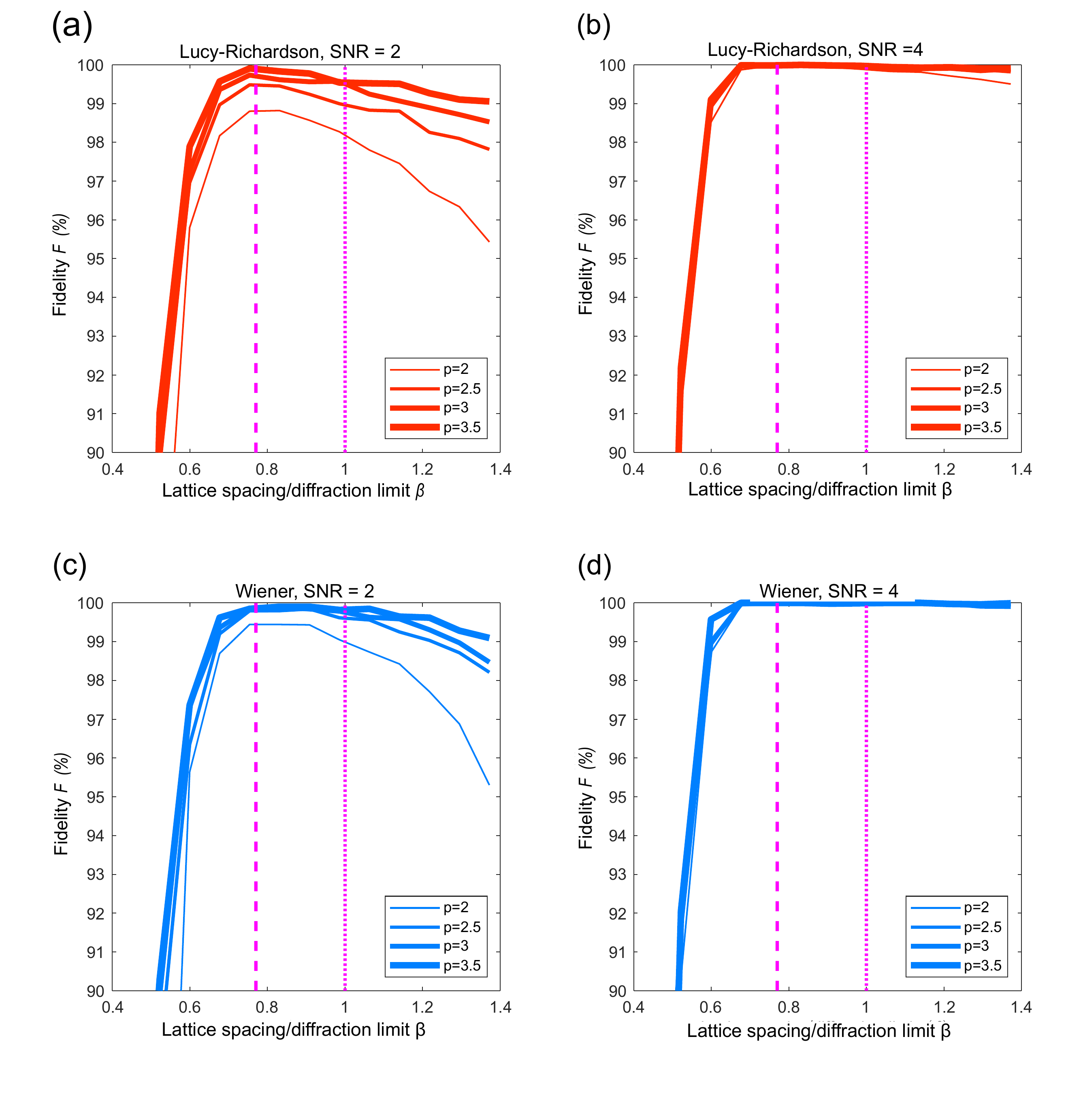} 
	\caption[resolution]{% 
 Reconstruction fidelity of simulated images with 50\% filling, for varying resolution parameters $\beta = a / d_{min}$, and pixelation $p=2.5, 3, 3.5$ and $4$. a) and b) The fidelity for the LR algorithm for images with 
 $ \SNR = 2$ and $\SNR=4$. Seven Lucy-Richardson iterations are used for all deconvolutions. c) and d) The fidelity for the Wiener algorithm with $\SNR =2$  and $\SNR=4$. Each point is an average fidelity from five simulated images.  The Rayleigh and Sparrow diffraction limits are shown as dotted and dashed vertical lines respectively.  
  		\label{fig:resolution} 
 }
\end{figure}

Here we investigate both the effect of changing the number of pixels per lattice site $p$ and the effect of operating near or below the diffraction limit independently. We generated simulated images where the lattice spacing is varied from $a=\unit[200]{nm}$ to $a=\unit[900]{nm}$. This corresponds to varying the ratio of lattice spacing over diffraction limit, $\beta = a / d_{min}$, in the range $0.29 < \beta < 1.37 $. We vary the pixelation in four steps from 2.5 to 4 pixels per lattice site while we keep the magnification fixed. When increasing the lattice spacing in an experiment one would automatically increase the pixelation $p$. To keep them independent the pixel size is adjusted to keep $p$ constant while $a$ is varied. The local-iterative deconvolution is no longer used here as it achieved consistently lower fidelities. We evaluate the reconstruction fidelity for $\SNR=2$ and $\SNR=4$ , and we see a markedly reduced fidelity for atom separations well below the Rayleigh limit for both deconvolution techniques (Fig.~\ref{fig:resolution}). The distance between objects where the local minimum in between their overlapping point-spread functions vanishes is known as Sparrow's limit. This limit at \unit[0.77]{$d_{min}$} matches the point at which perfect reconstruction becomes impossible. In Fig.~\ref{fig:resolution}, Rayleigh's and Sparrow's limits are shown by the dashed and dotted lines, respectively. For $\beta \geq 0.6 $ one can keep a high fidelity ($F > $\unit[97]{\%} ) even at SNR $=2$. 

When comparing the results for $\SNR=2$  and $\SNR=4$ , one finds that more pixels per lattice spacing can improve the fidelity, most notably when the signal is lower. The Nyquist sampling limit can be recognised in our results, because only for  $p>3.0$ we reach perfect reconstruction. We found that when the signal is equal to the background noise ($\text{SNR}=1$) there is no observable scaling with the lattice spacing (or $\beta$) and increasing the pixelation cannot increase the fidelity beyond 95\%. Increasing the signal over $\text{SNR}=4$ only leads to marginal improvements below the diffraction limit but does not move the steep decline in fidelity at $\beta=0.5$. Likewise, increasing the pixelation over $p=5$ does not lead to any further improvements in fidelity. 

Both the LR and Wiener protocols show a reduced fidelity when the lattice spacing is increased beyond the resolution limit, and when the signal is low. 
This counter-intuitive reduction occurs when increasing the lattice spacing with fixed $p$. As we keep the pixelation and lattice spacing independent in our simulations, there is a decrease in the fidelity when increasing the lattice spacing in the low signal regime ($\text{SNR}=2$). In this regime, more signal from an atom is detected on just a single pixel as we keep a fixed number of pixels per lattice spacing. As the lattice spacing is increased also the noise increases as each pixel receives more background light. While the SNR stays the same, it becomes harder to distinguish atoms whose signal is captured mostly by a single pixel from pixels with a high noise value. At even larger lattice spacings, $a > $ \unit[1500]{nm}, which we have not considered in this study, all the light from an  atom would be captured by a single pixel, at which point increasing the pixel would only add additional noise. Both LR and Wiener methods show a similar scaling of the fidelity with $\beta$ for $\text{SNR}= 2 \text{ and } 4 $. The fidelity only drops significantly with high $\beta$ when the pixelation is below the Nyquist sampling limit.
 
\section{Conclusion}
We have studied the detection fidelity of three different deconvolution techniques on simulated quantum-gas microscope images. The fidelity of each method scales differently with SNR level and lattice filling. The LR method was found to be the best which yielded the highest reconstruction fidelities. The LI method was found to be least efficient for high-filling ($\eta> 50 \% $) and required most computation time. The LR method was least affected by inhomogeneities in the atom brightness. Both the LR and Wiener deconvolution were able to accurately identify atoms on lattice sites separated by 0.6 times the diffraction limit, provided that the pixelation is above the Nyquist sampling threshold and that the images have a good signal-to-noise ($\text{SNR}>4.0$). It was shown how the reconstruction fidelities are affected when the signal decreases and when a lattice site is imaged on fewer pixels.
Our methods are scalable and applicable to a wide-range of quantum-gas microscope images with different resolution, wavelengths and lattice spacing. While many experiments using bosonic alkali atoms can benefit from a good signal of $\text{SNR} \geq 3$, imaging other atomic species is more challenging. When imaging fermionic alkali atoms we obtain at a typical SNR of $2$ \cite{haller2015single} and in new experiments with ultracold molecules and (non-cooled) earth-alkalis, even lower signal levels of SNR $\approx 1.5$ are encountered \cite{rosenberg2022observation, miranda2015site}.  
Using our methods, reliable atom detection far below the diffraction limit ($\beta < 0.5 $) is not possible.
However, several other experimental techniques have been used to achieve super-resolution of ultracold atoms in 1D trap geometries beyond the diffraction limit \cite{mcdonald2019superresolution,subhankar2019nanoscale,deist2022superresolution}.
Another promising approach for both 1D and 2D applications is  machine learning. A recent study \cite{picard2019deep} has shown that a trained neural network is able to detect atoms in  challenging signal-to-noise regimes, e.g., for non-cooled Ytterbium atoms \cite{miranda2017site}.  The challenge for machine-learning approaches is to choose a suitable  network architecture and to train it correctly and reliably \cite{picard2019deep, impertro2022unsupervised}.  By nature, both super-resolution techniques and the neural networks often depend on the specific experiment which makes it harder to make a one-to-one comparison to the techniques that are considered here. Ultimately, the reconstruction fidelity is limited by atom hopping and loss events that can occur during imaging. In future studies, these effects could also be incorporated in the simulation  \cite{haller2015single, hilker2017spin},  as well as possible sub- and super-radiance effects
\cite{facchinetti2016storing,facchinetti2018interaction,rui2020subradiant}, which cause a density-dependent fluorescence per atom.

\section{Acknowledgments} 
We acknowledge support by the EPSRC through the Programme Grant DesOEQ (EP/P009565/1), the Quantum Technology Hub in Quantum Computing and Simulation (EP/T001062/1) and the New Investigator Grant (EP/T027789/1).

\newpage

\section{References}

\bibliographystyle{iopart-num}

\bibliography{myreferences}

\providecommand{\newblock}{}
\begin{thebibliography}{10}
\expandafter\ifx\csname url\endcsname\relax
  \def\url#1{{\tt #1}}\fi
\expandafter\ifx\csname urlprefix\endcsname\relax\def\urlprefix{URL }\fi
\providecommand{\eprint}[2][]{\url{#2}}

\bibitem{bakr2009quantum}
Bakr W~S, Gillen J~I, Peng A, F{\"o}lling S and Greiner M 2009 {\em Nature\/}
  {\bf 462} 74--77

\bibitem{bloch2008many}
Bloch I, Dalibard J and Zwerger W 2008 {\em Rev. Mod. Phys.\/} {\bf 80} 885

\bibitem{sherson2010single}
Sherson J~F, Weitenberg C, Endres M, Cheneau M, Bloch I and Kuhr S 2010 {\em
  Nature\/} {\bf 467} 68--72

\bibitem{bakr2010probing}
Bakr W~S, Peng A, Tai M~E, Ma R, Simon J, Gillen J~I, Foelling S, Pollet L and
  Greiner M 2010 {\em Science\/} {\bf 329} 547--550

\bibitem{cheuk2015quantum}
Cheuk L~W, Nichols M~A, Okan M, Gersdorf T, Ramasesh V~V, Bakr W~S, Lompe T and
  Zwierlein M~W 2015 {\em Phys. Rev. Lett.\/} {\bf 114} 193001

\bibitem{greif2016site}
Greif D, Parsons M~F, Mazurenko A, Chiu C~S, Blatt S, Huber F, Ji G and Greiner
  M 2016 {\em Science\/} {\bf 351} 953--957

\bibitem{edge2015imaging}
Edge G~J, Anderson R, Jervis D, McKay D~C, Day R, Trotzky S and Thywissen J~H
  2015 {\em Phys. Rev. A\/} {\bf 92} 063406

\bibitem{boll2016spin}
Boll M, Hilker T~A, Salomon G, Omran A, Nespolo J, Pollet L, Bloch I and Gross
  C 2016 {\em Science\/} {\bf 353} 1257--1260

\bibitem{mazurenko2017cold}
Mazurenko A, Chiu C~S, Ji G, Parsons M~F, Kan{\'a}sz-Nagy M, Schmidt R, Grusdt
  F, Demler E, Greif D and Greiner M 2017 {\em Nature\/} {\bf 545} 462--466

\bibitem{brown2017spin}
Brown P~T, Mitra D, Guardado-Sanchez E, Schau{\ss} P, Kondov S~S, Khatami E,
  Paiva T, Trivedi N, Huse D~A and Bakr W~S 2017 {\em Science\/} {\bf 357}
  1385--1388

\bibitem{cheuk2016observation}
Cheuk L~W, Nichols M~A, Lawrence K~R, Okan M, Zhang H, Khatami E, Trivedi N,
  Paiva T, Rigol M and Zwierlein M~W 2016 {\em Science\/} {\bf 353} 1260--1264

\bibitem{simon2011quantum}
Simon J, Bakr W~S, Ma R, Tai M~E, Preiss P~M and Greiner M 2011 {\em Nature\/}
  {\bf 472} 307--312

\bibitem{fukuhara2013quantum}
Fukuhara T, Kantian A, Endres M, Cheneau M, Schau{\ss} P, Hild S, Bellem D,
  Schollw{\"o}ck U, Giamarchi T, Gross C {\em et~al.\/} 2013 {\em Nature
  Phys\/} {\bf 9} 235--241

\bibitem{preiss2015strongly}
Preiss P~M, Ma R, Tai M~E, Lukin A, Rispoli M, Zupancic P, Lahini Y, Islam R
  and Greiner M 2015 {\em Science\/} {\bf 347} 1229--1233

\bibitem{bohrdt2019classifying}
Bohrdt A, Chiu C~S, Ji G, Xu M, Greif D, Greiner M, Demler E, Grusdt F and Knap
  M 2019 {\em Nat. Phys.\/} {\bf 15} 921--924

\bibitem{kwon2022site}
Kwon K, Kim K, Hur J, Huh S and Choi J~Y 2022 {\em Phys. Rev. A\/} {\bf 105}
  033323

\bibitem{yamamoto2020single}
Yamamoto R, Ozawa H, Nak D~C, Nakamura I and Fukuhara T 2020 {\em New J.
  Phys.\/} {\bf 22} 123028

\bibitem{liu2021site}
Liu L, Mongkolkiattichai J, Yang J and Schauss P 2021 {\em PRX Quantum\/} {\bf
  2} 020344

\bibitem{asteria2021quantum}
Asteria L, Zahn H~P, Kosch M~N, Sengstock K and Weitenberg C 2021 {\em
  Nature\/} {\bf 599} 571--575

\bibitem{nelson2007imaging}
Nelson K~D, Li X and Weiss D~S 2007 {\em Nature Phys\/} {\bf 3} 556--560

\bibitem{krahn2021erbium}
Krahn A~J 2021 {\em Erbium Quantum Gas Microscope\/} Ph.D. thesis Harvard
  University

\bibitem{knottnerus2020microscope}
Knottnerus I, Pyatchenkov S, Onishchenko O, Urech A, Schreck F and Siviloglou G
  2020 {\em Opt. Express\/} {\bf 28} 11106--11116

\bibitem{picard2019deep}
Picard L~R, Mark M~J, Ferlaino F and van Bijnen R 2019 {\em Meas. Sci.
  Technol.\/} {\bf 31} 025201

\bibitem{yamamoto2016ytterbium}
Yamamoto R, Kobayashi J, Kuno T, Kato K and Takahashi Y 2016 {\em New J.
  Phys.\/} {\bf 18} 023016

\bibitem{phelps2019dipolar}
Phelps G~A 2019 {\em A dipolar quantum gas microscope\/} Ph.D. thesis
  Massachusetts Institute of Technology

\bibitem{lucy1974iterative}
Lucy L~B 1974 {\em Astronomical Journal\/} {\bf 79} 745

\bibitem{richardson1972bayesian}
Richardson W~H 1972 {\em JoSA\/} {\bf 62} 55--59

\bibitem{wiener1964extrapolation}
Wiener N 1964 {\em Extrapolation, Interpolation, and Smoothing of Stationary
  Time Series: With Engineering Applications\/} vol~8 (MIT Press)

\bibitem{alberti2016super}
Alberti A, Robens C, Alt W, Brakhane S, Karski M, Reimann R, Widera A and
  Meschede D 2016 {\em New J. Phys.\/} {\bf 18} 053010

\bibitem{mcdonald2019superresolution}
McDonald M, Trisnadi J, Yao K~X and Chin C 2019 {\em Physical Review X\/} {\bf
  9} 021001

\bibitem{subhankar2019nanoscale}
Subhankar S, Wang Y, Tsui T~C, Rolston S and Porto J~V 2019 {\em Physical
  Review X\/} {\bf 9} 021002

\bibitem{deist2022superresolution}
Deist E, Gerber J~A, Lu Y~H, Zeiher J and Stamper-Kurn D~M 2022 {\em Physical
  Review Letters\/} {\bf 128} 083201

\bibitem{haller2015single}
Haller E, Hudson J, Kelly A, Cotta D~A, Peaudecerf B, Bruce G~D and Kuhr S 2015
  {\em Nature Phys\/} {\bf 11} 738--742

\bibitem{andor}
{ANDOR iXon Ultra 897} (
  andor.oxinst.com/assets/uploads/products/andor/documents/andor-ixon-ultra-emccd-specifications.pdf
  )

\bibitem{MatlabPoisson}
Poissonian filter in matlab
  \urlprefix\url{https://uk.mathworks.com/help/stats/poissrnd.html}

\bibitem{manwar2021signal}
Manwar R, Zafar M and Xu Q 2021 {\em Optics\/} {\bf 2} 1--24

\bibitem{starck2002deconvolution}
Starck J~L, Pantin E and Murtagh F 2002 {\em PASP\/} {\bf 114} 1051

\bibitem{WienerMatlab}
{Wiener deconvolution in Matlab}
  \urlprefix\url{https://uk.mathworks.com/help/images/deblurring-images-using-a-wiener-filter.html}

\bibitem{LucyMatlab}
{Lucy Richardson deconvolution in Matlab}
  \urlprefix\url{https://uk.mathworks.com/help/images/ref/deconvlucy.html}

\bibitem{schauss2015high}
Schauss P 2015 {\em High-resolution imaging of ordering in Rydberg many-body
  systems\/} Ph.D. thesis LMU M{\"u}nchen

\bibitem{MatlabImresize}
Image resize in matlab
  \urlprefix\url{https://uk.mathworks.com/help/matlab/ref/imresize.html}

\bibitem{ockeloen2010detection}
Ockeloen C, Tauschinsky A, Spreeuw R and Whitlock S 2010 {\em Physical Review
  A\/} {\bf 82} 061606

\bibitem{niu2018optimized}
Niu L, Guo X, Zhan Y, Chen X, Liu W and Zhou X 2018 {\em Applied Physics
  Letters\/} {\bf 113} 144103

\bibitem{la2018deposition}
La~Rooij A, Couet S, Van Der~Krogt M, Vantomme A, Temst K and Spreeuw R 2018
  {\em Journal of Applied Physics\/} {\bf 124} 044902

\bibitem{gonzalez2015subwavelength}
Gonz{\'a}lez-Tudela A, Hung C~L, Chang D~E, Cirac J~I and Kimble H 2015 {\em
  Nature Photon\/} {\bf 9} 320--325

\bibitem{wang2018dark}
Wang Y, Subhankar S, Bienias P, {\L}{\k{a}}cki M, Tsui T~C, Baranov M~A,
  Gorshkov A~V, Zoller P, Porto J~V and Rolston S~L 2018 {\em Phys. Rev.
  Lett.\/} {\bf 120} 083601

\bibitem{pawley2006points}
Pawley J~B 2006 Points, pixels, and gray levels: Digitizing image data {\em
  Handbook of Biological Confocal Microscopy\/} (Springer) pp 59--79

\bibitem{rosenberg2022observation}
Rosenberg J~S, Christakis L, Guardado-Sanchez E, Yan Z~Z and Bakr W~S 2022 {\em
  Nature Phys\/} {\bf 18} 1062--1066

\bibitem{miranda2015site}
Miranda M, Inoue R, Okuyama Y, Nakamoto A and Kozuma M 2015 {\em Physical
  Review A\/} {\bf 91} 063414

\bibitem{miranda2017site}
Miranda M, Inoue R, Tambo N and Kozuma M 2017 {\em Physical Review A\/} {\bf
  96} 043626

\bibitem{impertro2022unsupervised}
Impertro A, Wienand J~F, H{\"a}fele S, von Raven H, Hubele S, Klostermann T,
  Cabrera C~R, Bloch I and Aidelsburger M 2022 {\em arXiv preprint
  arXiv:2212.11974\/}

\bibitem{hilker2017spin}
Hilker T~A 2017 {\em Spin-resolved microscopy of strongly correlated fermionic
  many-body states\/} Ph.D. thesis LMU M{\"u}nchen

\bibitem{facchinetti2016storing}
Facchinetti G, Jenkins S~D and Ruostekoski J 2016 {\em Phys. Rev. Lett.\/} {\bf
  117} 243601

\bibitem{facchinetti2018interaction}
Facchinetti G and Ruostekoski J 2018 {\em Phys. Rev. A\/} {\bf 97} 023833

\bibitem{rui2020subradiant}
Rui J, Wei D, Rubio-Abadal A, Hollerith S, Zeiher J, Stamper-Kurn D~M, Gross C
  and Bloch I 2020 {\em Nature\/} {\bf 583} 369--374

\end{thebibliography}

\end{document}